\documentclass{PoS}
\usepackage{psfig,epsfig}
\newcommand\npb[3]     {{\it Nucl.\ Phys.\ }{\bf B #1} (#2) #3}

\newcommand\prl[3]     {{\it Phys.\ Rev.\ Lett.\ }{\bf #1} (#2) #3}
\newcommand{\hepth}[1] {{\tt hep-th/#1}}
\newcommand\jhep[3]    {{\it J. High Energy Phys.\ }{\bf #1} (#2) #3}
\newcommand{\heplat}[1]{{\tt hep-lat/#1}}
\newcommand\prep[3]    {{\it Phys.\ Rept.\ }{\bf #1} (#2) #3}
\newcommand\plb[3]     {{\it Phys.\ Lett.\ }{\bf B #1} (#2) #3}

\newcommand\prd[3]     {{\it Phys.\ Rev.\ }{\bf D #1} (#2) #3}

\title{
The spectrum of SU($N$) gauge theories at $\theta\ne 0$
}

\ShortTitle{SU($N$) spectrum at $\theta\ne 0$}

\author{Luigi Del Debbio \\
        SUPA, School of Physics, University of Edinburgh, 
        Edinburgh EH9 3JZ, UK\\
        E-mail: \email{luigi.del.debbio@ed.ac.uk} 
} 

\author{Gian Mario Manca \\
        Dipartimento di Fisica dell'Universit\`a di Parma and I.N.F.N., 
        I-43100 Parma, Italy \\ 
        E-mail: \email{manca@fis.unipr.it} 
} 

\author{\speaker{Haralambos Panagopoulos} \\ 
        Department of Physics, University of Cyprus,
        Lefkosia, CY-1678, Cyprus\\ 
        E-mail: \email{haris@ucy.ac.cy} 
}

\author{Apostolos Skouroupathis\thanks{Supported in part by the 
Research Promotion Foundation of Cyprus} \\ 
        Department of Physics, University of Cyprus,
        Lefkosia, CY-1678, Cyprus\\ 
        E-mail: \email{php4as01@ucy.ac.cy} 
}

\author{Ettore Vicari \\ 
        Dipartimento di Fisica dell'Universit\`a 
        di Pisa and I.N.F.N., I-56127 Pisa, Italy \\ 
        E-mail: \email{vicari@df.unipi.it} 
}

\abstract{ We study the $\theta$ dependence of the spectrum of
  four-dimensional SU($N$) gauge theories, where $\theta$ is the coefficient
  of the topological term in the Lagrangian, for $N\ge 3$ and in the large-$N$
  limit.  We compute the $O(\theta^2)$ terms of the expansions around
  $\theta=0$ of the string tension and the lowest glueball mass, respectively
  $\sigma(\theta) = \sigma \left( 1 + s_2 \theta^2 + ... \right)$ and
  $M(\theta) = M \left( 1 + g_2 \theta^2 + ... \right)$, where $\sigma$ and
  $M$ are the values at $\theta=0$.  For this purpose we use numerical
  simulations of the Wilson lattice formulation of SU($N$) gauge theories for
  $N=3,4,6$.  The $O(\theta^2)$ coefficients turn out to be very small for all
  $N\ge 3$.  For example, $s_2=-0.08(1)$ and $g_2=-0.06(2)$ for $N=3$.  Their
  absolute values decrease with increasing $N$.  Our results are suggestive of
  a scenario in which the $\theta$ dependence in the string and glueball
  spectrum vanishes in the large-$N$ limit, at least for sufficiently small
  values of $|\theta|$. They support the general
  large-$N$ scaling arguments that indicate $\bar{\theta}\equiv \theta/N$ as
  the relevant Lagrangian parameter in the large-$N$ expansion.  }

\FullConference{XXIV International Symposium on Lattice Field
Theory\\
         July 23-28 2006\\
         Tucson Arizona, US}

\begin{document}

\section{Introduction}
\label{intro}
Four-dimensional SU($N$) gauge theories have a nontrivial dependence
on the angle $\theta$ that appears in the Euclidean Lagrangian as
\begin{equation}
{\cal L}_\theta  = {1\over 4} F_{\mu\nu}^a(x)F_{\mu\nu}^a(x)
- i \theta {g^2\over 64\pi^2} \epsilon_{\mu\nu\rho\sigma}
F_{\mu\nu}^a(x) F_{\rho\sigma}^a(x)
\label{lagrangian}
\end{equation}
($q(x)=\frac{g^2}{64\pi^2} \epsilon_{\mu\nu\rho\sigma}
F_{\mu\nu}^a(x) F_{\rho\sigma}^a(x)$ is the topological
charge density).
Indeed, the most plausible explanation of how the solution of the
so-called U(1)$_A$ problem can be compatible with the $1/N$ expansion
requires a nontrivial
$\theta$ dependence of the ground-state energy density $F(\theta)$, 
\begin{equation}
F(\theta) = -{1\over V} \ln \int [dA] \exp \left(  - \int d^d x {\cal L}_\theta
\right),
\label{GSE}
\end{equation}
to leading order in $1/N$. 
The large-$N$ ground-state energy is expected to behave as
\cite{Witten-98,Gabadadze}
\begin{equation}
\Delta F(\theta)\equiv F(\theta) - F(0) = {\cal A} \, \theta^2  +
O\left( 1/N^2\right) 
\label{conj}
\end{equation} 
for suf{f}iciently small $\theta$, i.e. $\theta<\pi$.  This
has been supported by Monte Carlo simulations \cite{DPV-02}. 
Indeed, numerical results for $N=3,4,6$ are consistent with
a scaling behavior around $\theta=0$\,:
\begin{equation}
f(\theta) \equiv \sigma^{-2} \Delta F(\theta) 
= {1\over 2} C \theta^2 ( 1 + b_2 \theta^2 + ...), \quad
C=C_\infty + c_2/N^2 + ... , \quad b_2=b_{2,2}/N^2+...
\label{ftheta}
\end{equation}
where $\sigma$ is the string tension at $\theta=0$.
$C$ is the ratio
$\chi/\sigma^2$ where $\chi = \int d^4 x \langle q(x)q(0) \rangle$ 
is the topological susceptibility at $\theta=0$.
Its large-$N$ limit $C_\infty$ is \cite{DPV-02,LT-01}
$C_{\infty}\approx 0.022$. Estimates of 
$c_2$ and $b_{2,2}$ are \cite{DPV-02} $c_2\approx
0.06$, and $b_{2,2}\approx -0.2$
($b_2\approx -0.02$ for SU(3)\,).
Eq.~(\ref{ftheta}) can be recast as
\begin{equation} 
f(\theta) = N^2 \bar{f}(\bar{\theta}\equiv \theta/N), 
\qquad\bar{f}(\bar{\theta}) = 
{1\over 2} C \bar{\theta}^2 ( 1 + \bar{b}_2 \bar{\theta}^2 + ...), 
\label{fthetabar}
\end{equation}
where $\bar{b}_2=b_{2,2}+ O(1/N^2)=O(1)$.  This is consistent with
$1/N$ scaling arguments, indicating $\bar{\theta}\equiv \theta/N$ as
the relevant Lagrangian 
parameter in the large-$N$ limit of the ground-state
energy.

The $\theta$ dependence of the
spectrum is particularly interesting in the
large-$N$ limit, which may also be addressed by other approaches,
such as AdS/CFT correspondence, see e.g. Ref.~\cite{AGMOO-00}. The
analysis  of glueballs using AdS/CFT
suggests that the only effect of $\theta$ in the leading
large-$N$ limit is that the lowest glueball state becomes a
mixed $0^{++}/0^{-+}$ state, but its mass does not
change \cite{Gabadadze}.

We present an exploratory study of $\theta$
dependence in the spectrum of SU($N$) gauge theories, using
numerical simulations of the Wilson action. MC
studies are very dif{f}icult in the presence of the complex
valued $\theta$ term, defying simulation. 
We focus on relatively small
$\theta$, where one may expand observable values about $\theta=0$.
For the string tension and the lowest glueball mass we write
\begin{equation}
\sigma(\theta) = \sigma \left( 1 + s_2 \theta^2 + ... \right),\qquad
M(\theta) = M\left( 1 + g_2 \theta^2 + ... \right)
\label{gmex}
\end{equation} 
where $M$ is the $0^{++}$ glueball mass at $\theta=0$.  The
dimensionless quantities $s_2$ and $g_2$ can be computed from
correlators at $\theta=0$; they should
approach constants as $a\to 0$, with $O(a^2)$ corrections.  

We present results for 4-d SU($N$) gauge theories with
$N=3,4,6$.  The $O(\theta^2)$ coef{f}icients turn out
very small for all $N\ge 3$, e.g., $s_2=-0.08(1)$ and $g_2=-0.06(2)$
for $N=3$.  Moreover, their absolute values decrease with $N$. \ 
$O(\theta^2)$ terms are substantially smaller in
dimensionless ratios such as $M/\sqrt{\sigma}$ and ratios of
independent $k$ strings, $R_k=\sigma_k/\sigma$.  Our results suggest
a large-$N$ scenario in which $\theta$ dependence in the string and
glueball spectrum vanishes around $\theta=0$. They are consistent with
arguments indicating $\bar{\theta}\equiv \theta/N$
as the relevant parameter in the large-$N$ limit.  We also show a similar
situation in the 2-d CP$^{N-1}$ models by an analysis of
their $1/N$ expansion.

In Sec.~\ref{sec2} we outline the
method to estimate $O(\theta^2)$ terms of the expansion
in $\theta$. The results of our
numerical study are presented in Sec.~\ref{res}.
In Sec.~\ref{cpn} we discuss the $\theta$ dependence of
2-d CP$^{N-1}$ models. Ref.~\cite{DMPSV} is a longer write-up of this
work, with a more complete list of references.

\section{Numerical method}
\label{sec2}

\subsection{Monte Carlo simulations}
\label{mc}

In our simulations of lattice gauge theories, using the Wilson
formulation, we employed the 
Cabibbo-Marinari algorithm 
to upgrade SU($N$) matrices by updating their SU(2)
subgroups. This was done by alternating
microcanonical over-relaxation and heat-bath steps, typically in a 4:1
ratio. 

The topological properties of {f}ields de{f}ined on a lattice
are strictly trivial.
Physical topological properties are recovered
in the continuum limit.  Various techniques have been proposed to
associate a topological charge $Q$ to a lattice con{f}iguration;
the most robust de{f}inition of $Q$ uses the index
of the overlap Dirac operator. However, due to the computational cost 
of fermionic methods and the need for very large statistics to measure 
correlations of Polyakov and plaquette operators with topological quantities,
we used the simpler cooling method, implemented as in Ref.~\cite{DPV-02}. 
Comparison with a fermionic estimator shows good agreement
for SU(3)~\cite{DPV-02,DGP-05}.  Moreover,
the agreement among different methods is expected to improve for larger $N$.

A severe form of critical slowing down affects the measurement of $Q$, posing
a serious limitation for numerical studies, especially at large $N$.  
The available estimates of the autocorrelation time $\tau_Q$ for
topological modes appear to increase
as an exponential or a large power of the length scale
\cite{DPV-02,DMV-04}.
This dramatic effect has not
been observed in plaquette-plaquette or Polyakov line correlations, suggesting
an approximate decoupling between topological and nontopological ones,
such as those determining con{f}ining properties and the glueball spectrum.
But, as we shall see, such a decoupling is not complete. Therefore the strong
critical slowing down that is observed in the topological sector will
eventually affect also the measurements of nontopological quantities.

\subsection{The $O(\theta^2)$ coef{f}icients of the $\theta$ expansion}
\label{thex}

Let us describe how to determine the 
$O(\theta^2)$ coef{f}icients in Eq.~(\ref{gmex}).
We {f}irst discuss the string tension; it
can be determined from the torelon mass, i.e. the
mass describing the large-time exponential 
decay of wall-wall correlations $G_P$ of Polyakov lines.  
In the presence of a $\theta$ term
\begin{equation}
G_P(t,\theta)=\langle A_P(t) \rangle_\theta = 
{ \int [dU] A_P(t) e^{-\int d^4 x {\cal L}_\theta}
\over \int [dU] e^{-\int d^4 x {\cal L}_\theta }}\ ,
\qquad
A_P(t) = \sum_{x_1,x_2}  {\rm Tr}\,P(0;0) \; {\rm Tr}\,P(x_1,x_2;t),  
\label{apdef}
\end{equation}
$P(x_1,x_2;t)$ is the Polyakov line of
size $L$ along the $x_3$ direction.
The time separation $t$ is an integer multiple of the lattice spacing
$a$: $t = n_t\,a$.
The correlation $G_P$ can be expanded in $\theta$\,: 
\begin{equation}
G_P(t,\theta) = G_P^{(0)}(t) + \frac{1}{2}\theta^2 G_P^{(2)}(t) + O(\theta^4),
\label{g2}
\end{equation}
where:\ \  $
G_P^{(0)}(t) = \langle A_P(t) \rangle_{\theta=0},\qquad
G_P^{(2)}(t) = - \langle A_P(t) Q^2 \rangle_{\theta=0} + 
\langle A_P(t) \rangle_{\theta=0} \langle Q^2 \rangle_{\theta=0}$\,.

The correlation function $G_P$ is expected to have
a large-$t$ exponential behavior
\begin{equation}
G_P(t,\theta) \approx B(\theta) e^{-E(\theta) t},
\label{gptlarge}
\end{equation}
where  $E(\theta)$ is the energy of the lowest state, and $B(\theta)$
is the overlap of the  
source with this state.
If $L$ is suf{f}iciently large, the lowest-energy states
should be those of a string-like spectrum.
\begin{equation}
E(\theta) = \sigma(\theta) L - {\pi/ (3L)} 
\label{lute}
\end{equation}

We expand the large-$t$ behavior (\ref{gptlarge}) of $G(t,\theta)$ as
\begin{equation}
G_P(t,\theta) \approx B_0 e^{- E_0 t} \left[ 1 + \theta^2 h(t) + ... \right]
\label{explt}
\end{equation}
where:\ \ 
$B(\theta) = B_0 + \theta^2 B_2 + ... ,\qquad
E(\theta) = E_{0} + \theta^2 E_{2} + ...,\qquad
h(t) = ({B_2}/{B_0}) - E_{2} t \,. $

Comparing Eq.~(\ref{explt}) with Eq.~(\ref{g2}), 
we {f}ind that
\begin{equation}
h(t) = {G^{(2)}_2(t) \,/\, (2 G^{(0)}(t)) }
\end{equation}
Thus $E_2$ can be estimated from the difference:
$\Delta h(t) = h(t)-h(t+a).$ Indeed,
${\displaystyle \lim_{t\to\infty}} \; \Delta h(t) = E_2\, a$\,.
Corrections are exponentially suppressed 
as $\exp[-(E^*_0-E_0) t]$, where $E^*_0$ is the mass of the {f}irst excited
state at $\theta=0$.  Assuming the free-string spectrum, 
$E^*_0 - E_0= 4\pi/L$.  Since we choose the lattice size
$L$ so that $l_\sigma \equiv \sqrt{\sigma} L \approx 3$, 
$(E_0^*-E_0)/E_0 \approx 4\pi/l_\sigma^2 \approx 1.4$.

{F}inally, the dimensionless scaling coef{f}icient $s_2$ of the
$O(\theta^2)$ term in (\ref{gmex}) is obtained by
\begin{equation}
s_2= {E_{2}\over \sigma L}
\label{s2ext}
\end{equation}
$s_2$ is expected to approach a constant in the continuum limit, with
$O(a^2)$ 
scaling corrections.

An analogous procedure can be used for the lowest $0^{++}$
glueball 
mass $M(\theta)$.  We
employ wall-wall correlators of Wilson loops with up to 6 spatial
links. We de{f}ine:
$\Delta k(t) \equiv k(t)-k(t+a)$, 
where $k(t)$ is analogous to $h(t)$, using glueball correlators. 
Then, $g_2$ in (\ref{gmex})
is obtained by 
\begin{equation}
g_2 = {1\over a\, M} \,\lim_{t\to\infty} \; \Delta k(t) 
\label{g2ext}
\end{equation}

In order to improve the ef{f}iciency of the
measurements we used smearing and blocking procedures to construct
operators with better overlaps. Our implementation is
described in Ref.~\cite{DPRV-02}. 

\section{Results}
\label{res}
\TABLE[h]{
\caption{ 
Information on our MC simulations.
Estimates of $\sigma$ are obtained
using Eq. (\protect\ref{lute}).
}
\label{tableruns}
\footnotesize
\begin{tabular}{cccclll}
\hline\hline
\multicolumn{1}{c}{$N$}&
\multicolumn{1}{c}{$\beta$}&
\multicolumn{1}{c}{lattice}&
\multicolumn{1}{c}{stat}&
\multicolumn{1}{c}{$a^2\,\sigma$}&
\multicolumn{1}{c}{$a\,M_{0^{++}}$}&
\multicolumn{1}{c}{$M_{0^{++}}/\sqrt{\sigma}$}\\
\hline \hline
3 & 5.9 & $12^3\times 18$ & 25M/20 & 0.0664(6)  & 0.80(1) & 3.09(4) \\

3 & 6.0 & $16^3\times 36$  & 25M/40 & 0.0470(3) & 0.70(1) & 3.23(4) \\

4 & 10.85 & $12^3\times 18$ & 16M/50 & 0.0646(6)& 0.76(1) & 2.99(5) \\

6 & 24.5  & $8^3\times 12$  & 9M/50 & 0.114(2)  & 0.83(1) & 2.46(4) \\

\hline\hline
\end{tabular}
}
Table~\ref{tableruns} contains some information on our MC runs for
$N=3,4,6$ on lattices $L^3\times T$.  Since the coef{f}icients of the $\theta$
expansions are computed from connected
correlation functions, and turn out to be quite small,
high statistics is required to distinguish their estimates from zero: Our runs
range from 9 to 25 million sweeps, with measures taken every 20-50 sweeps.
This requirement represents a serious limitation to the possibility of
performing runs for large lattices and in the continuum limit, especially for
large $N$, due also to the severe critical slowing down. For all
values of $\beta$ considered, $\tau_Q$
satis{f}ies $\tau_Q \lesssim 100$ ~\cite{DPV-02}.
Furthermore, $\beta$ values were chosen in the weak-coupling
region (see Ref.~\cite{DPRV-02} for a more detailed discussion of this
point). The lattice size $L$
was chosen so that $l_\sigma\equiv \sqrt{\sigma}L\gtrsim 3$ (see, e.g.,
Refs.~\cite{LT-01,DPRV-02}).  Due to these
limitations, in particular for $N=4,6$, we could afford
only one value of $\beta$, so that no stringent checks of scaling could be
performed. For this reason our study should be still considered as a {f}irst
exploratory investigation.

\vskip 0.3cm
\noindent
\begin{minipage}{0.48\linewidth}
\begin{center}
\epsfig{file=n3b5p9.eps, width=7truecm}
\end{center}
\end{minipage}\hskip0.04\textwidth
\begin{minipage}{0.48\linewidth}
\begin{center}
\epsfig{file=n3b6.eps, width=7truecm}
\end{center}
\end{minipage}

\smallskip\centerline{{\small{{\bf {F}igure 1}: 
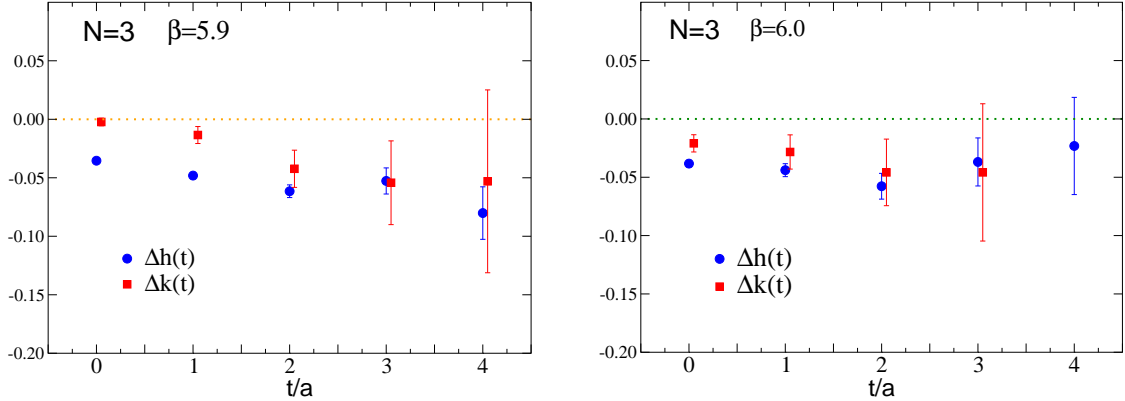
Plot of $\Delta h(t)$ and $\Delta k(t)$ for $N=3$ at $\beta=5.9$ 
and $\beta=6.0$\,.
}}}

\medskip
{F}igs.~1 and 2 show the results for the discrete differences
$\Delta h(t)$ and $\Delta k(t)$ for
$N=3$ at $\beta=5.9,6.0$ and for $N=4,6$ respectively.  As
expected, the signal degrades rapidly with increasing $t$.  Anyway, they
appear rather stable already for small values of $t$. 
In the case $N=3$ the data at $\beta=6$
appear to approach the asymptotic behavior more rapidly than at $\beta=5.9$.
This should be due to the fact that more effective blocking can be
applied when $L=16$, rather than $L=12$.

We estimate $s_2$ and
$g_2$ in (\ref{gmex}) from $\Delta h(t)$ and
$\Delta k(t)$,
taking the data at $t/a=2$ in the $N=3,4$ runs, and at $t/a=1$ for $N=6$.  In
Table~\ref{tableres} we report the results.  The estimates of $s_2$ and $g_2$
are small in all cases, and decrease with increasing $N$.  For $N=3$ the
results at $\beta=5.9$ and $\beta=6.0$ are consistent, supporting the expected
scaling behavior.  As {f}inal estimate one may consider
\begin{equation}
s_2=-0.08(1), \quad g_2=-0.06(2) \qquad {\rm for}\;\; N=3
\end{equation}

One may also consider the the scaling ratio:
${M(\theta)/ \sqrt{\sigma(\theta)}} =
({M / \sqrt{\sigma}}) ( 1 +  c_2 \theta^2 + ... )$,
where $c_2=g_2-s_2/2$. Using the results of Table~\ref{tableres},
we see that the $O(\theta^2)$ terms tend to cancel in the ratio.
Indeed, we {f}ind $c_2=-0.02(2),\, -0.01(3),\, -0.01(2)$ respectively 
for $N=3,4,6$. 

For $N>3$ there are additional
independent $k$-strings associated with representations
of higher $n$-ality. 
One may consider the ratio $R_k(\theta) = \sigma_k(\theta)/\sigma(\theta)
= R_k ( 1 + r_{k,2} \,\theta^2 + ... )$,
where $\sigma_k$ is the $k$-string tension (see e.g.
Refs.~\cite{LT-01,DPRV-02,DPV-03,LTW-04}). For $N=4$ there is
one additional $k$ string, $\sigma_2$\,; for
$N=6$ there are two.  Our results for $k>1$ strings are less stable.
We obtained suf{f}iciently precise results only for $N=4$.
They suggest a very weak $\theta$-dependence in $R_2$,
i.e. $|r_{2,2}|\lesssim 0.02$. 

\vskip 0.3cm
\noindent
\begin{minipage}{0.48\linewidth}
\begin{center}
\epsfig{file=n4b.eps, width=7truecm}
\end{center}
\end{minipage}\hskip0.04\textwidth
\begin{minipage}{0.48\linewidth}
\begin{center}
\epsfig{file=n6b.eps, width=7truecm}
\end{center}
\end{minipage}

\smallskip\centerline{{\small{{\bf {F}igure 2}: 
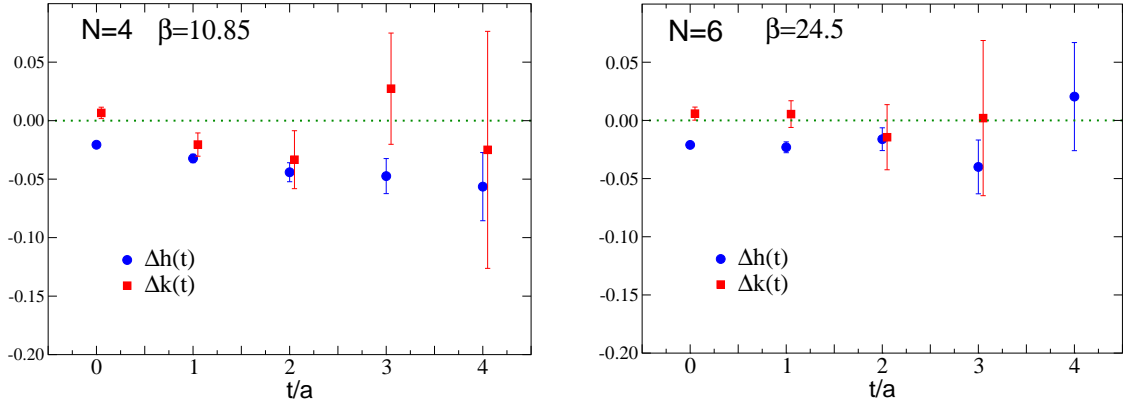
Plot of $\Delta h(t)$ and $\Delta k(t)$ for
$N=4$ at $\beta=10.85$ and $N=6$ at $\beta=24.5$\,.
}}}

\hfill
\TABLE[ht]{
\caption{
Results for $s_2$ and $g_2$, as derived from the discrete
differences at $t/a=2$ for $N=3,4$, and at $t/a=1$ for $N=6$.  
}
\label{tableres}
\footnotesize
\begin{tabular}{ccccll}
\hline\hline
\multicolumn{1}{c}{$N$}&&
\multicolumn{1}{c}{$\beta$}&&
\multicolumn{1}{c}{$s_2$}&
\multicolumn{1}{c}{$g_2$}\\
\hline \hline
3 &\qquad\qquad& 5.9  &\qquad\qquad&  $-$0.077(8)\qquad\qquad   & $-$0.05(2) \\

3 && 6.0  &&  $-$0.077(15)  & $-$0.07(4) \\

4 && 10.85 &&  $-$0.057(10) &  $-$0.04(3) \\

6 && 24.5  &&  $-$0.025(5) & $\phantom{-}$0.006(15) \\

\hline\hline
\end{tabular}
}\hfill

\noindent
In conclusion, the above results show that $O(\theta^2)$ terms in the
spectrum of SU($N$) gauge theories are very
small, especially in dimensionless ratios.  Moreover, they
decrease with increasing $N$, and the coef{f}icients do not appear
to converge to a nonzero value.  This suggests a scenario
in which the $\theta$ dependence of the spectrum disappears at large
$N$. 
General arguments indicate $\bar{\theta}\equiv
\theta/N$ as the relevant parameter, implying that
$O(\theta^2)$ coef{f}icients in the spectrum should decrease 
as $1/N^{2}$.  This is roughly veri{f}ied by our results, 
barring possible scaling corrections, especially for
$N=4,6$. For example, in the case of the string tension,
$s_2\approx s_{2,2}/N^2$ with $s_{2,2}\approx -0.9$\,.
 Of course, further investigations are required to put this scenario
on a {f}irmer ground.

Recent
studies~\cite{DPV-04,LTW-05} at {f}inite temperature have shown that in
the large-$N$ limit the 
topological properties remain substantially unchanged up to the {f}irst-order
transition point.

\section{$\theta$ dependence in the two-dimensional CP$^{N-1}$ model}
\label{cpn}

Issues concerning the $\theta$ dependence can also be discussed in
two-dimensional CP$^{N-1}$ models~\cite{DDL-79}, which 
present several
features of QCD and, in addition, are amenable to a systematic $1/N$
expansion around the large-$N$ saddle-point
solution~\cite{DDL-79,CR}.

One may expand the ground state energy $F(\theta)$ about $\theta=0$.
De{f}ining a scaling quantity
$f(\theta)$:
\begin{equation}
f(\theta) \equiv M^{-2} [F(\theta)-F(0)] 
= {1\over 2} C \theta^2 ( 1 + \sum_{n=1} b_{2n} \theta^{2n} ) 
\label{fthetacpn} 
\end{equation}
$M$ is the ``zero momentum'' mass
at $\theta=0$, and $C$ is the
ratio $\chi/M^2$ at $\theta=0$, where $\chi$ is the topological
susceptibility. Within the $1/N$ expansion: 
$C = \chi/M^2 = {1/ (2\pi N)} + O(1/N^2)$. 
One obtains $b_{2n}$ from 
correlation functions of $q(x)$.
The analysis of the $1/N$-expansion Feynman diagrams
shows that $b_{2n}$ is suppressed as:
$b_{2n} = O(1/N^{2n})$. 
Thus the ground-state energy becomes
\allowbreak
\begin{equation} 
f(\theta) = N \bar{f}(\bar{\theta}\equiv \theta/N), 
\qquad\bar{f}(\bar{\theta}) = 
{1\over 2} \bar{C} \bar{\theta}^2 
( 1 + \sum_{n=1} \bar{b}_{2n} \bar{\theta}^{2n} ), 
\label{fthetabarcpn}\end{equation}
\allowbreak
where $\bar{C}\equiv N C$ and $\bar{b}_{2n}=N^{2n}b_{2n}$ are $O(1)$ in
the large-$N$ limit.  Note the analogy with SU($N$) gauge theories.
The calculation of $b_{2n}$ is rather cumbersome; we obtain
\begin{equation}
\bar{b}_2= - {27/ 5}, \qquad \bar{b}_4= -{1830/ 7}.
\end{equation}

Within the $1/N$ expansion one may also study the dependence of
$M$ on $\theta$. We write: 
$M(\theta) = M\left( 1 + m_2 \theta^2 + ... \right)$.
A diagrammatic analysis indicates that $m_2$ is suppressed as:
$m_2 = O(1/N^2)$.
This con{f}irms the arguments indicating
$\bar{\theta}\equiv \theta/N$ as the relevant parameter in the
large-$N$ limit.

\end{document}